\documentclass[aps,prc,groupedaddress,twocolumn,showpacs,amsmath,amssymb]{revtex4}
\usepackage{graphics}% Include figure files
\usepackage{dcolumn}% Align table columns on decimal point 
\usepackage{longtable}

\begin{document}

\title{Consistent optical potential for incident and emitted low-energy $\alpha$ particles. II. $\alpha$-emission in fast-neutron induced reactions on Zr isotopes}
\author{V.~Avrigeanu} \email{vlad.avrigeanu@nipne.ro}
\author{M.~Avrigeanu}
\affiliation{Horia Hulubei National Institute for Physics and Nuclear Engineering, P.O. Box MG-6, 077125 Bucharest-Magurele, Romania}
%\date{\today}

\begin{abstract}
%\begin{description} 
\noindent
{\bf Background:} Challenging questions of the $\alpha$-particle optical-model potential (OMP)
are still pointed out by recent high-precision measurements of $\alpha$-induced reaction data below the Coulomb barrier. Moreover, the reliability of a previous OMP for $\alpha$-particles on nuclei within the mass number range 45$\leq$$A$$\leq$209 has been recently proved for emitted $\alpha$ particles as well, but only in the case of proton-induced reactions on Zn isotopes [Phys. Rev. C {\bf 91}, 064611 (2015), Paper I]. 

\smallskip

\noindent
{\bf Purpose:} 
Analysis of most recent $(\alpha,\gamma)$ reaction data for Ge and Zr isotopes, which provides an additional  validation of the above-mentioned potential, is related to a further account of $\alpha$-particle emission in neutron-induced reactions on Zr isotopes, at the same time with a suitable description of all competitive processes.  

\smallskip

\noindent
{\bf Methods:} A consistent parameter set, established or validated by independent analysis of  recent various data, particularly $\gamma$-ray strength functions, have been involved within  model calculation of the $(\alpha,\gamma)$ as well as $(n,\alpha)$ reaction cross sections. The latter are part of the whole analysis of the neutron activation of Zr isotopes, in order to avoid any error compensation or latent ambiguity. 

\smallskip

\noindent
{\bf Results:} 
The aforesaid potential provides a consistent description of recent $\alpha$-induced reaction data with no empirical rescaling factors of the $\gamma$ and/or nucleon widths. On the other hand, its use leads to underestimated predictions of the pre-equilibrium emission and statistical models for the $(n,\alpha)$ reaction cross sections. 

\smallskip

\noindent
{\bf Conclusions}: 
An optical potential with a volume imaginary component seems to be needed to describe the low-energy $\alpha$-particle evaporation, while only surface absorption occurs in $\alpha$-induced reactions at similar energies. 
%\end{description}
\end{abstract}


\maketitle

\section{Introduction}
\label{intro}
The recently improved cross sections for reactions induced by reaction-in-flight (RIF) neutrons with energies up to 30 MeV in warm deuterium-tritium plasma \cite{bc16}, have concerned also $^{94}$Zr$(n,\alpha)^{91}$Sr reaction. 
A significant body of measured data becomes thus available for $\alpha$-particle emission in fast-neutron induced reactions on Zr isotopes, to be eventually used for assessment of the so-called $\alpha$-potential mystery for the account at once of both absorption and emission of low-energy $\alpha$-particles \cite{tr13,va15,va160}. 

Actually, it is yet open the question of a need for new physics in potentials to describe particle evaporation from a transient nuclear stratosphere of the emitter nucleus. 
The  density of such an excited nucleus should differ from cold nuclei (\cite{rl87} and Refs. therein) and be thus particularly considered within a microscopic optical model potential (OMP) formalism \cite{ma06}.
Meanwhile, an optical potential providing a suitable description of the incident $\alpha$-particle data within the mass range 45$\leq$$A$$\leq$209 \cite{va14,va16} has been proved to describe also the $\alpha$-emission in low-energy proton-induced reactions on Zn isotopes. Nevertheless, it has been found necessary to carry on additional similar analyses over the same $A$ range \cite{va15} (Paper I). 
Besides its basic interest, an accurate description of the $\alpha$-particle OMP is highly required by many nuclear astrophysics applications and estimation of radiation damage effects. 

On the other hand, a latest extensive study of fast-neutron induced reaction on Zr isotopes above 15 MeV \cite{vs10} has shown significant problems to describe particularly the $(n,p)$ and $(n,\alpha)$ reactions. 
Thus, it is almost definitely a challenge the suitable account of both the absolute cross sections and excitation-function trends of these reactions in the energy range of interest for H and He production and related radiation damage calculations.
In fact, a first goal of Ref. \cite{vs10} was to reuse as many as possible of the parameters optimized in the framework of assessing the available fast-neutron reaction data for $^{89}$Y, versus globally optimized phenomenological model parameters as well as globally optimized microscopic calculations. 
The phenomenological models for direct interaction (DI), pre-equilibrium emission (PE) and compound-nucleus (CN) statistical equilibration of an excited nucleus were involved in this respect within version 1.0 of the computer code TALYS \cite{ajk12,TALYS}. 
The agreement with the data was considered good only in view of the lack of detailed tuning of calculations, while especially with reference to $\alpha$-particle emission it was stated the need of detailed further investigations for the improved physics modeling. Therefore, in order to obtain reliable results of the $(n,\alpha)$ reaction analysis for Zr isotopes and the related $\alpha$-particle OMP, first it has become necessary a check of the model predictions for the rest of the measured fast-neutron reaction data available for these isotopes \cite{bc16,vs10,exfor}.

As a matter of fact, since the beginning of the modern nuclear data evaluation (e.g., \cite{eda80}) the mass region $A$$\sim$90 was considered to provide unusual conditions under which nuclear models and parameters can be tested. 
This could explain the presence of the above-mentioned problems somehow at variance with the usual lack of critical deficiencies in statistical-model (SM) calculations \cite{tk15}. 
Moreover, recent studies of nucleon- \cite{sg03,pt04,sh13,ln14} and especially $\alpha$-capture \cite{sjq15} pointed out in this mass region more problems in using different combinations of SM parameters for a consistent description of either both $(p,\gamma)$ and $(p,n)$ channels data for $^{89}$Y target nucleus \cite{sh13} or $(\alpha,\gamma)$ reaction on Zr isotopes \cite{sjq15}.
This issue has firstly been essential even for the validation of the above-mentioned $\alpha$-particle OMP. 
Furthermore, about half of the measured fast-neutron reaction data for Zr isotopes being  isomeric cross sections, the account of the $\gamma$-decay of the corresponding excited nuclei is quite important for their model calculations. 

Ultimately, to obtain reliable conclusions on the $\alpha$-capture and $(n,\alpha)$ reaction assessment for Zr isotopes and the related $\alpha$-particle OMP, a consistent analysis (e.g., \cite{eda80,ma88}) has concerned in the present work (i) the same common parameters being used within the corresponding OMP, PE, and SM  models, (ii) the use of a consistent input parameter set, either established or validated by analyzing various independent data, and (iii) the simultaneous {\bf not a fit but model account} of the available neutron-reaction data for all stable Zr isotopes using the same parameter set and no empirical rescaling factors of the $\gamma$ and/or nucleon widths. Unphysical calculations or parameters resulting from the cross-section analysis of a single reaction can be thus largely avoided \cite{eda80b}.

The models and parameters involved in the present work are briefly mentioned in Sec.~\ref{SMcalc}.   
Validation of the $\alpha$-particle OMP \cite{va14} through the analysis of the new $(\alpha,\gamma)$ reaction cross section is proved in Sec.~\ref{AlphaCap}.
The results obtained for the fast-neutron induced reactions on Zr stable isotopes are then compared with the measured data \cite{bc16,vs10,exfor} in Sec.~\ref{Res}, while the case of the $\alpha$-particle emission makes the object of Sec.~\ref{aOMP}. Conclusions are finally given in Sec.~\ref{Conc}.
Preliminary results were described elsewhere  \cite{ND2016va}.

\begingroup
\squeezetable
\begin{table*} % add [H] placement to break table across pages
\caption{\label{densp} Low-lying levels number $N_d$ up to excitation energy $E^*_d$ \protect\cite{ensdf} used in SM calculations of reaction cross sections, the low-lying levels and $s$-wave nucleon-resonance spacings $D_0^{\it exp}$ (with uncertainties given in parentheses, in units of the last digit) in the energy range $\Delta$$E$ above the separation energy $S$, for the target-nucleus ground state (g.s.) spin $I_0$, fitted to obtain the BSFG level-density parameter {\it a} and g.s. shift $\Delta$ (for a spin cutoff factor calculated with a variable moment of inertia \cite{va02} between half and 75\% of the rigid-body value, from g.s. to $S$, and reduced radius $r_0$=1.25 fm).} 
\begin{ruledtabular}
\begin{tabular}{cccccccccr} 
Nucleus   &$N_d$&$E^*_d$& \multicolumn{5}{c}
                  {Fitted level and resonance data}& $a$ & $\Delta$\hspace*{3mm}\\
\cline{4-8}
           &  &     &$N_d$&$E^*_d$&$S+\frac{\Delta E}{2}$&
                                     $I_0$&$D_0^{\it exp}$ \\ 
           &  &(MeV)&   & (MeV)& (MeV)&  &(keV)&(MeV$^{-1}$) & (MeV) \\ 
\noalign{\smallskip}\hline\noalign{\smallskip}
$^{73}$Se&27&1.092&27&1.092& 8.532 & 0 & 0.32(12) & 9.60&-1.37 \\
$^{74}$Se&26&2.919&26&2.919&       &   &          & 9.50& 0.54 \\ 
$^{77}$Se&33&1.282&33&1.282& 7.426 & 0 & 0.65(10) &10.02&-1.18 \\
$^{78}$Se&33&2.949&33&2.949&10.498 &1/2& 0.120(15)& 9.88& 0.48 \\ 

$^{84}$Rb&22&1.007&22&1.007&       &   &          &10.00&-1.27 \\ 
$^{85}$Rb&26&1.950&26&1.950&       &   &          &10.00&-0.37 \\ 
$^{86}$Rb&24&1.559&24&1.559& 8.661 &5/2& 0.172(8) & 9.21&-0.96 \\ 

$^{84}$Sr&25&3.332&25&3.332&       &   &          & 9.60& 0.95 \\ 
$^{85}$Sr&27&1.712&33&1.850& 8.532 & 0 & 0.32(12) &10.64&-0.46 \\
$^{86}$Sr&21&3.186&19&3.104&       &   &          & 9.30& 0.79 \\ 
$^{87}$Sr&20&2.539&29&2.708& 8.442 & 0 &  2.6(8)  & 9.12& 0.04 \\ 
$^{88}$Sr&33&4.515&47&4.801&11.113 &9/2& 0.29(8)  & 8.70& 1.63 \\
$^{89}$Sr&28&3.433&22&3.249& 6.430 & 0 & 23.7(29) & 9.58& 0.87 \\  
$^{90}$Sr&15&3.039&17&3.146&       &   &          & 9.60& 0.95 \\
$^{91}$Sr&11&1.942&11&1.942&       &   &          &10.00& 0.09 \\  
$^{92}$Sr&15&2.925&33&4.614&       &   &          &10.00& 0.89 \\
$^{93}$Sr&22&2.292&20&2.169&       &   &          &10.60& 0.09 \\  
$^{94}$Sr&23&3.155&23&3.155&       &   &          &11.00& 1.08 \\

$^{86}$ Y&21&1.277&21&1.277&       &   &          & 9.40&-1.12 \\ 
$^{87}$ Y&24&1.849&64&2.502&       &   &          & 9.50&-0.57 \\ 
$^{88}$ Y&24&1.477&17&1.262&       &   &          & 9.40&-1.12 \\ 
$^{89}$ Y&26&3.630&26&3.630&11.478 & 4 &0.106(35)$^a$&8.90&0.94\\ 
$^{90}$ Y&17&1.815&18&1.962& 6.857 &1/2&  3.7(4)  & 9.18&-0.38 \\ 
$^{91}$ Y&11&1.580&10&1.547&       &   &          & 9.30&-0.40 \\ 
$^{92}$ Y& 4&0.431& 4&0.431&       &   &          &10.40&-1.00 \\ 
$^{93}$ Y&22&2.200&21&2.129&       &   &          &10.10&-0.08 \\ 
$^{94}$ Y& 4&0.724&[4&0.431]$^b$&  &   &          &11.40&-0.80 \\ 
$^{95}$ Y&10&2.047&10&2.047&       &   &          &11.40& 0.50 \\ 
$^{96}$ Y& 3&0.652&[4&0.431]$^b$&  &   &          &12.00&-0.70 \\ 

%$^{87}$Zr&24&1.949&24&1.949&       &   &          & 9.15&-0.58 \\ 
$^{88}$Zr&27&3.484&27&3.484&       &   &          & 8.75& 0.75 \\ 
$^{89}$Zr&30&2.572&30&2.572&       &   &          & 9.20&-0.10 \\ 
$^{90}$Zr&41&4.701&41&4.701&       &   &          & 9.00& 1.73 \\ 
$^{91}$Zr&32&2.928&32&2.928& 7.260 & 0 & 6.0(14)  & 9.70& 0.35 \\ 
$^{92}$Zr&42&3.500&54&3.725& 8.647 &5/2& 0.55(10) & 9.65& 0.77 \\ 
$^{93}$Zr&21&2.095&21&2.095& 6.785 & 0 & 3.5(8)   &10.50&-0.02 \\ 
$^{94}$Zr&23&3.059&23&3.059& 8.220 &5/2& 0.302(75)&10.96& 1.00 \\ 
$^{95}$Zr&20&2.372&20&2.372& 6.507 & 0 & 4.0(8)   &11.31& 0.44 \\ 
$^{96}$Zr&38&3.630&38&3.630&       &   &          &11.20& 1.32 \\ 
$^{97}$Zr& 4&1.400& 4&1.400& 5.629 & 0 & 13(3)    &11.40& 0.30 \\

$^{93}$Mo&58&2.915&58&2.915& 8.092 & 0 & 2.7(5)   & 9.35&-0.18 \\ 
$^{94}$Mo&54&3.401&60&3.462& 9.678 &5/2&0.081(24)$^c$&10.74& 0.78 \\ 
$^{95}$Mo&27&1.692&27&1.692& 7.377 & 0 & 1.32(18) &10.40&-0.61 \\ 
$^{96}$Mo&38&2.875&38&2.875& 9.154 &5/2&0.661(30)$^c$&11.35&0.61 \\ 

\end{tabular}	 
\end{ruledtabular}
\begin{flushleft}
$^a$Reference \cite{mg14}\\ 
$^b$Levels of $^{92}$Y nucleus\\
$^c$Reference \cite{hu13}\\ 
\end{flushleft}
\end{table*}
\endgroup

\section{Nuclear models and parameters}
\label{SMcalc}

The SM Hauser-Feshbach (HF) \cite{wh52} and PE Geometry-Dependent Hybrid (GDH) \cite{mb83} model calculations were carried out in this work using an updated version of the computer code STAPRE-H95 \cite{ma95}, including the OM code SCAT2 \cite{scat2} as a subroutine.
Values of $\sim$0.2 and 0.4 MeV equidistant binning were used for the excitation energy grid for either capture or particle-emission analysis. 

The DI distorted-wave Born approximation (DWBA) method and a local version of the computer code DWUCK4 \cite{pdk84} were also used for calculation of the collective inelastic scattering cross sections. These results were then involved for the subsequent decrease of the total-reaction cross section $\sigma_R$ within the PE+HF calculations. 

The corresponding results, obtained within a local approach as shown in the following, are also compared with calculated reaction cross sections provided by use of the code TALYS-1.8 and its default input parameters \cite{TALYS}. The content of the evaluated data library TENDL-2015 \cite{TENDL15} which is based on particularly adjusted TALYS calculations in order to describe the measured data, has been used in the same respect, too. 

A consistent set of (i) back-shifted Fermi gas (BSFG) \cite{hv88} nuclear level densities, (ii) nucleon and (iii) $\gamma$-ray transmission coefficients was used also within the present analysis of the fast-neutron induced reactions on the stable isotopes of Zr. 
These parameters were established or validated on the basis of independent low-lying levels \cite{ensdf} and nucleon resonance data \cite{ripl3}, $(p,n)$ reaction cross sections \cite{exfor}, $\gamma$-ray strength functions \cite{gs79,acl16,gmt16,OCL} and $(p,\gamma)$ reaction cross sections \cite{exfor}, respectively. 
The same OMP and level density parameters have been used in the framework of the DI, PE, and SM models. 
Only points in addition to the details given formerly \cite{va15,ma06,va14,va16,va02,pr05,ma09,ma09b,ma10,ma12,ma13} as well as the particular parameter values are mentioned hereafter.

\subsection{Nuclear level densities}
\label{NLD}

The BSFG model parameters used in this work are given in Table~\ref{densp} at once with the low-lying level numbers and excitation energies \protect\cite{ensdf} used either at once in the SM calculations (the 2nd and 3rd columns) or formerly, along with the resonance data, in their setting up. Nuclei in addition to those concerned previously within Ref. \cite{ma13} are included in this table as well as BSFG-parameter updates. These updates, particularly for $^{89}$Y and $^{90}$Zr semi-magic nuclei, concern changes between 0.6--5\% for the level density parameter $a$ in order to fit, together with the corresponding changes of the g.s. back-shift $\Delta$, the low-lying levels and resonance data. 
The smooth-curve method \cite{chj77} was applied for nuclei without resonance data, using average $a$-values for the $A$$\sim$90 and the $\Delta$ values obtained alone by fit of the low-lying discrete levels. 
These changes followed either the availability of new data published in the meantime \cite{ensdf}, or the increased attention paid to the accurate account of Y and Zr isotopes which have been now of larger interest than previously \cite{ma13}. 

A note should concern the level-density spin distribution determined by a variable ratio $I/I_r$ of the nuclear moment of inertia to its rigid-body value,  i.e., between  0.5 for  ground states, 0.75 at the neutron binding energy, and 1 around the excitation energy of 15 MeV \cite{va02}. It is quite important for the model calculations of isomeric cross sections (e.g., \cite{ma12} and Refs. therein).  
The fact that the variable ratio $I/I_r$ corresponds to suitable $\sigma^2_d$ values in the energy range of the discrete levels \cite{ajk08,tve09,mag85}, close to the assumption of Koning {\it et al.} \cite{ajk08} at the neutron binding energy, and in agreement with theoretical predictions (\cite{djd95} and Refs. therein) at higher energies, is shown for $A$$\sim$90 in Fig. 6 of Ref. \cite{ma13}.

\subsection{Optical model potentials} \label{OMP}

{\it The neutron optical-potential} local parameters of Koning and Delaroche \cite{KD03} for Y and Zr isotopes were adopted to obtain the transmission coefficients for neutrons. 
These potentials were used also for the DWBA calculation of the DI collective inelastic scattering cross sections, using the corresponding deformation parameters of the first 2$^+$ and 3$^-$ collective states \cite{ensdf}. The weak coupling model was adopted for the odd nucleus $^{91}$Zr using also the collective state parameters of Kalbach \cite{ck00}.
Typical DI inelastic-scattering cross sections decrease from 5--6 to 3--4\% in the energy range from few to $\sim$22 MeV.
This approach should be involved prior to model calculations with PE+SM codes as STAPRE-H, while it is built-in within the complex code TALYS. 

\begin{figure} %[b]
\resizebox{0.75\columnwidth}{!}{\includegraphics{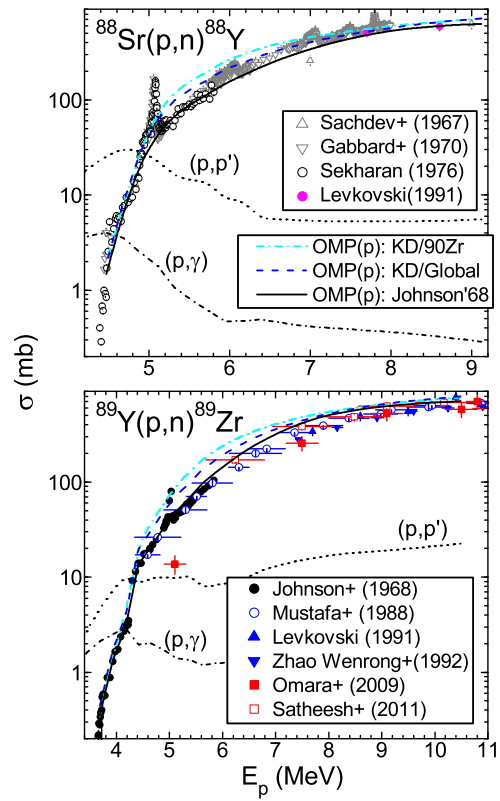}}
\caption{\label{fig:SrYpn}(Color online) Comparison of the measured cross sections \cite{exfor} of the $(p,n)$ reaction on $^{88}$Sr (top) and $^{89}$Y (bottom) and calculated values corresponding to proton OMP parameters of either the global (dashed curves) or local set for $^{90}$Zr (dash-dotted curves) of Ref. \cite{KD03}, and that of Johnson {\it et al.} \cite{chj68} (solid curves). The finally calculated cross sections of $(p,p')$ reaction (short-dashed curves) and $(p,\gamma)$ reaction (short dash-dotted curves) are also shown.}
\end{figure}

{\it The proton optical potential} of Koning and Delaroche \cite{KD03} was also the first option for calculation of the proton transmission coefficients for Sr and Y residual nuclei in neutron-induced reactions on Zr isotopes. However, an overestimation was proved for the use of this OMP within detailed analysis of the $(p,\gamma)$ and $(p,n)$ reactions on $^{88}$Sr \cite{sg03} and $^{89}$Y \cite{pt04,sh13} target nuclei at astrophysically relevant energies. 
Actually the authors of Ref. \cite{KD03} noted that their predicted proton reaction cross sections are slightly higher than most of the fitted data points for $^{90}$Zr, but a suitable OMP assessment was provided by the corresponding differential data. However all these data are at energies above the range of interest of the present work, while it is already known that the absorption cross section is somewhat smaller than predicted by an OMP which fits proton elastic scattering at higher energies \cite{chj68}.
Therefore we carried out also an analysis of these $(p,n)$ reaction cross sections up to several MeV above their effective thresholds (Fig.~\ref{fig:SrYpn}) while a distinct discussion in Sec.~\ref{RSF} concerns the related $(p,\gamma)$ reactions. 

The corresponding SM calculations were obviously carried out using the same input parameters as in the rest of this work.  
Thus, we found that the measured  excitation functions \cite{exfor} are overestimated, especially in the first 2-3 MeV, with $\sim$45\% by the global parameters of Ref. \cite{KD03} and even $\sim$80\% by their local parameter set for protons on $^{90}$Zr. The former effect is also present in the case of the evaluated cross sections of TENDL-2015 library \cite{TENDL15}. 

This overestimation has been removed (Fig.~\ref{fig:SrYpn}) by using in this work the local OMP of Johnson {\it et al.} \cite{chj68} for low-energy protons on $^{89}$Y. 
A suitable account has thus been obtained for the entire off-resonance excitation functions while the data for the $d_{5/2}$ isobaric analog resonances observed at 5.07 MeV in the $^{88}$Sr$(p,n)^{88}$Y reaction, and 4.8 and 5.0 MeV in the $^{89}$Y$(p,n)^{89}$Zr reaction, were omitted from the fit performed to obtain the OMP parameters \cite{chj79}. Actually, the SM results were used in Ref. \cite{chj68} for subtraction of the non-resonant background and interpretation of the resonances themselves, whereas the correctness of the OMP parameters is of interest for the present work. 

The same potential has been used also for the heavier Sr and Y isotopes with the only change of the surface imaginary potential depth $W_D$ by taking into account its systematic but anomalous mass dependence \cite{chj79,dsf79}.  

{\it The $\alpha$-particle optical potential} for nuclei within the 45$\leq$$A$$\leq$209 range \cite{va14} was firstly used for both $\alpha$-induced reaction and $\alpha$-emission calculations, following the conclusions corresponding to the $A$$\sim$60 nuclei \cite{va15}. Moreover, this potential has recently been found to describe well the $(\alpha,n)$ reaction cross section for $^{84,86,87}$Sr isotopes \cite{ao16}. However, following the discussion on the $(n,\alpha)$ reactions in Sec.~\ref{aOMP}, the earlier OMP \cite{va94}, which was proved suitable for description of the $\alpha$-particle emission, has been used within present analysis too.

\subsection{$\gamma$-ray strength functions} \label{RSF}

We have continued to avoid the renormalization of $\gamma$-ray strength functions in order to achieve agreement between the measured and calculated capture cross sections but to rely \cite{va15,va14,va16} on the measured data of radiative strength function (RSF) and average $s$-wave radiation widths $\Gamma_{\gamma}$ \cite{ripl3}. 
The following comments concern the Y and Zr isotopes, involved in the neutron-induced reaction analysis, while the particular issues related to excited nuclei within the $(\alpha,\gamma)$ reaction are discussed in Sec.~\ref{AlphaCap} at once with the corresponding cross-section results.

\begin{figure} %[t]
\resizebox{0.75\columnwidth}{!}{\includegraphics{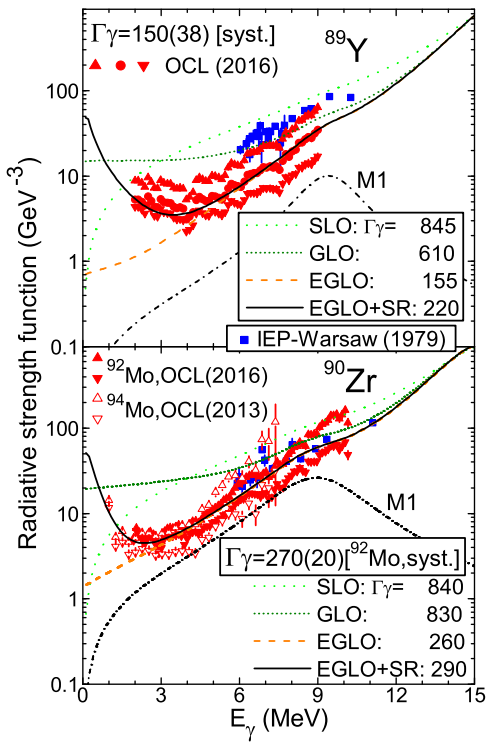}}
\caption{\label{fig:RSF_Y89Zr90}(Color online) Comparison of measured \cite{gs79,acl16,OCL} and calculated sum of $\gamma$-ray strength functions of the $E1$ and $M1$ radiations for $^{89}$Y (top) and $^{90}$Zr (bottom),  using the models SLO (dotted curves), GLO (short-dotted curves), and EGLO without (dashed curves) and with (solid curves) a SR contribution, for E1 radiations, and the SLO model for M1 radiations (short dash-dotted curves).  There are shown also measured dipole $\gamma$-ray strength functions for $^{92,94}$Mo nuclei \cite{gmt16,OCL} and $s$-wave average radiation widths $\Gamma_{\gamma}$ (in meV) either deduced from systematics for $^{89}$Y and $^{92}$Mo, or corresponding to M1 and each of above-mentioned E1 functions.} 
\end{figure}

There have been used in this respect the formerly measured RSFs for $^{89}$Y and $^{90}$Zr nuclei \cite{gs79} and especially the more recent high-accuracy measurements at lower energies for $^{89}$Y \cite{acl16}. There are shown in Fig.~\ref{fig:RSF_Y89Zr90} also the higher and lower limits of the quite recent data for $^{92,94}$Mo nuclei \cite{gmt16,OCL} because they are obviously rather similar to the former RSF data for $^{90}$Zr \cite{gs79}. 

Thus, the electric-dipole $\gamma$-ray strength functions, most important for calculation of the $\gamma$-ray  transmission coefficients, have been described by using the models of the former Lorentzian (SLO) \cite{pa62}, the generalized Lorentzian (GLO) \cite{jk91}, and finally the enhanced generalized Lorentzian (EGLO) \cite{jk93} with a constant nuclear temperature $T_f$ of the final states \cite{acl10}. 
The giant dipole resonance (GDR) line-shape usual parameters and the $T_f$ values from the studies for $^{89}$Y \cite{acl16}, $^{92}$Mo \cite{gmt16}, and $^{93-98}$Mo \cite{acl10}  nuclei were involved also in this work, as well as their SLO model parameters for the M1 radiation. 

A different choice has concerned however a small resonance (SR) of M1 type lastly used for $^{89}$Y \cite{acl16} to get a reasonable agreement with the measured strength. A similar SR  has been used in this work to describe the low-energy enhancement of the RSF data \cite{OCL}.  
The SR energy, width, and peak cross section of 0.6 MeV, 2.8 MeV, and 0.14 mb, respectively, for $^{89}$Y, and 0.4 MeV, 1.2 MeV, and 0.14 mb, for $^{90}$Zr, have actually been used as the $E1$ pigmy resonance of the GLO original formalism \cite{jk91} and provided at last a suitable account of the low-energy RSF data \cite{OCL} shown in Fig.~\ref{fig:RSF_Y89Zr90}.

The comparison of the measured and calculated sum of $\gamma$-ray strength functions of the $E1$ and $M1$ radiations for the nuclei $^{89}$Y and $^{90}$Zr shows that, similarly to other mass ranges \cite{va15,va14,va16}, both the SLO and GLO models lead to overestimation of the RSF data below the nucleon separation energy $S$. An  image of this overestimation is provided by the calculated $s$-wave average radiation widths $\Gamma_{\gamma}$ corresponding to the above-mentioned E1 and M1 models, which are also shown in Fig.~\ref{fig:RSF_Y89Zr90}. They are compared to the values either deduced from systematics of the measured-data dependence on the neutron $S$ (e.g., \cite{acl16,gmt16}), including the case of $^{92}$Mo nucleus with a similar nuclear structure to $^{90}$Zr \cite{gmt16}. Thus one may see that only the EGLO+SR $\gamma$-ray strength functions provide values closer to the measured data eventually in the limit of $2\sigma$ uncertainty, while the SLO and GLO models led to calculated values several times larger.  

One the other hand, it is obvious that the omission of the SR low-energy upbend contribution has, also within $2\sigma$ experimental uncertainty, less significant effects. Another spin-off result of the present work concerns the completion of a previous statement on the significant contribution of the RSF upbend to $\Gamma_{\gamma}$ for nuclei with small $S$ values \cite{gmt16}. As one may expect, it is confirmed now that this contribution depends also on the odd–-even character of the nucleus, as follows from the case of the semi-magic nuclei $^{89}$Y and $^{90}$Zr, with quite close $S$ values but with a SR contribution of $\sim$30\% for the odd nucleus $^{89}$Y but only $\sim$10\% for the even-even nucleus $^{90}$Zr.   

\begin{figure} %[b]
\resizebox{0.75\columnwidth}{!}{\includegraphics{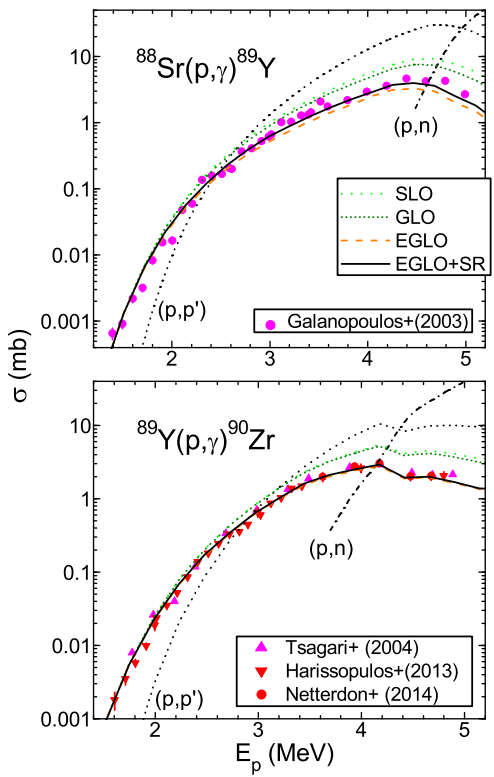}}
\caption{\label{fig:SrYpg}(Color online) Comparison of cross sections measured for $(p,\gamma)$ reaction on $^{88}$Sr \cite{sg03} (top) and $^{89}$Y \cite{pt04,sh13,ln14} (bottom) and calculated by using the E1 radiation RSF models SLO (dotted curves), GLO (short-dotted curves), EGLO (dashed curves), and EGLO+SR (solid curves). The finally calculated cross sections of $(p,p')$ reaction (short-dashed curves) and $(p,n)$ reaction (short dash-dotted curves) are also shown.}
\end{figure}

{\it $(p,\gamma)$ reaction data analysis} for the target nuclei $^{88}$Sr and $^{89}$Y (Fig.~\ref{fig:SrYpg}), has additionally been used to check the RSF accuracy. 
The comparison of these experimental and calculated capture cross sections is firstly pointing out a good agreement for the use of the EGLO+SR models. However, the SR contribution is not significant for the even-even residual nucleus $^{90}$Zr but increased for the odd nucleus $^{89}$Y. 

Moreover, one may note an increase by even a factor $\sim$2 of the calculated capture cross-section if the EGLO+SR model is replaced by the GLO one. A similar change but from GLO to SLO is followed by a much smaller increase, due to the $(p,p')$ and $(p,n)$ channels which become dominant at less than 1 MeV above their effective thresholds.  
Nevertheless, the present work shows more exactly than formerly \cite{va15,va16} that the low-energy RSF enhancement does affect the calculated $(p,\gamma)$ reaction cross sections much less than the use of either SLO or even GLO models. 

As a matter of fact, these results point out the usefulness of a consistent input parameters in nuclear model calculations, established or validated by analysis of various independent data, against the trial of different combinations of SM ingredients \cite{sg03,pt04} even self-consistent in the sense that they combine only phenomenological or microscopic models \cite{sh13}. Thus it has become obvious why, e.g., the OMP of Ref. \cite{KD03} leads to the $(p,\gamma)$ data overproduction that extends beyond two standard deviation \cite{sg03}. 
Moreover, only the present model calculations have been able to achieve the goal \cite{sh13} of a consistent description, by a single combination of SM ingredients, of both $(p,\gamma)$ and $(p,n)$ channels data also in the overlapping energy region from 3.6 to 5.2 MeV. 

\subsection{Pre-equilibrium emission modeling} \label{PE}

The PE contribution to the results of the present work is provided by the GDH model \cite{mb83}, which was generalized through inclusion of the angular-momentum and parity conservation \cite{ma88} and $\alpha$-particle emission based on a pre-formation probability $\varphi$ with the value 0.2 \cite{eg81}. It includes also a revised version of the advanced particle-hole level densities (PLD) \cite{ma94,ma98,ah98} using the Fermi-gas energy dependence of the single-particle level density \cite{ck85}. The particular energy dependence of the PE contribution within this approach is discussed at large in Sec. III.B.5 of Ref. \cite{pr05} for neutron-induced reactions on Mo isotopes. That discussion is thus fully appropriate also to this work. 

An additional note may concern the use of the central-well Fermi energy value $F$=40 MeV, while the local-density Fermi energies corresponding to various partial waves (e.g., Fig. 4 of Ref. \cite{pr05}) were provided within the local density approximation by the same OMP parameters given in Sec.~\ref{OMP}. Under these conditions, the PE fraction varies in the incident energy range 5-21 MeV from 2 to 30\% for $^{90}$Zr, from 6 to 36\% for $^{91}$Zr, from 4 to 31\% for $^{92}$Zr, from 5 to 36\% for $^{94}$Zr, and from 6 to 37\% for $^{96}$Zr. 

\begin{figure*} %[t]
\resizebox{1.5\columnwidth}{!}{\includegraphics{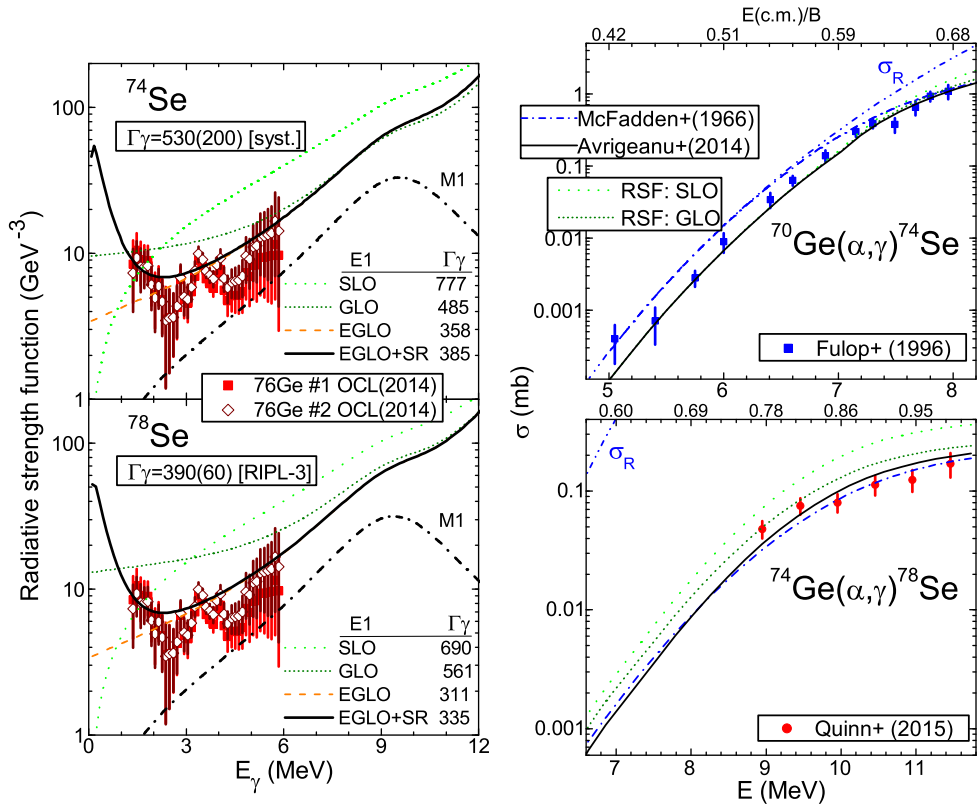}}
\caption{\label{fig:RSFag-Ge}(Color online) (left) As Fig.~\ref{fig:RSF_Y89Zr90} but for $^{74,78}$Se and dipole $\gamma$-ray strength functions for $^{76}$Ge nucleus \cite{as14,OCL}. (right) Comparison of measured cross sections of $(\alpha,\gamma)$ reaction on $^{70,74}$Ge \cite{sjq15,zf96} and calculated values using the $\alpha$-particle OMPs of Refs. \cite{mcf66} (dash-dotted curves) and \cite{va14} (solid curves), and alternate involvements for the latter OMP of either the GLO (short-dotted curves) and SLO (dotted curves) RSF models, vs. $\alpha$-particle laboratory energy (bottom) and ratio of center-of-mass energy to Coulomb barrier (top). There are shown also the $\sigma_R$ values given by  OMP \cite{mcf66} (dash-dot-dotted curves).} 
\end{figure*}

\section{Recent $(\alpha,\gamma)$ reaction data analysis} \label{AlphaCap}

A detailed study of the $(\alpha,\gamma)$ reactions on $^{74}$Ge and $^{90,92}$Zr nuclei \cite{sjq15} has been the newest issue of a major effort to provide a constraint for the choice of input models in a given $A$ range. As for all stable nickel isotopes \cite{as15}, different best combinations of input parameters for the TALYS 1.6 code were found for each of the investigated isotopes and also at variance with the grounds of the concerned $\alpha$-particle OMP \cite{pd02}. 

While the $(\alpha,\gamma)$ reaction data for $^{90,92}$Zr is of straightforward interest for the present work, the similar discussion for $^{74}$Ge completes the previous analysis of RSFs, $\alpha$-particle OMP, and $(\alpha,\gamma)$ reaction data for $A$$\sim$60 nuclei \cite{va15,va16} and the present $A$$\sim$90 ones. 
Moreover, while the corresponding RSF discussion is rather similar to that given in Sec.~\ref{RSF}, it is closely related in the following to that of $(\alpha,\gamma)$ cross sections. 
At the same time, the similar data already available for $^{70}$Ge \cite{zf96} and $^{91}$Zr \cite{sh05} are considered too, for a systematic analysis.

\subsection{$^{70,74}$Ge$(\alpha,\gamma)$$^{74,78}$Se} \label{Ge}

Besides the details given in Sec.~\ref{RSF} it may be noted that the adopted RSFs shown in Fig.~\ref{fig:RSFag-Ge} for $^{74,78}$Se have been obtained using the GDR parameters  derived from photoabsorption data for $^{78}$Se \cite{ripl3gamma}, the $T_f$=0.7 MeV value \cite{va15,va16}, the SR parameters given above for $^{90}$Zr, and the global parametrization \cite{ripl3} of the SLO model for the M1 radiation. The RSF calculated values are close to the high-accuracy data measured at lower energies for the neighboring even-even nucleus $^{76}$Ge \cite{as14}. Moreover, their agreement with the more recent measurements for $^{73,74}$Ge \cite{tr16} seems to be even better.

All remarks on the RSFs of $^{89}$Y and $^{90}$Zr in Sec.~\ref{RSF} are appropriate for $^{74,78}$Se as well. This includes the even more reduced SR contribution of the RSF low-energy upbend  to $\Gamma_{\gamma}$ values, of $\sim$7\%. 

\begin{figure*} %[b]
\resizebox{1.5\columnwidth}{!}{\includegraphics{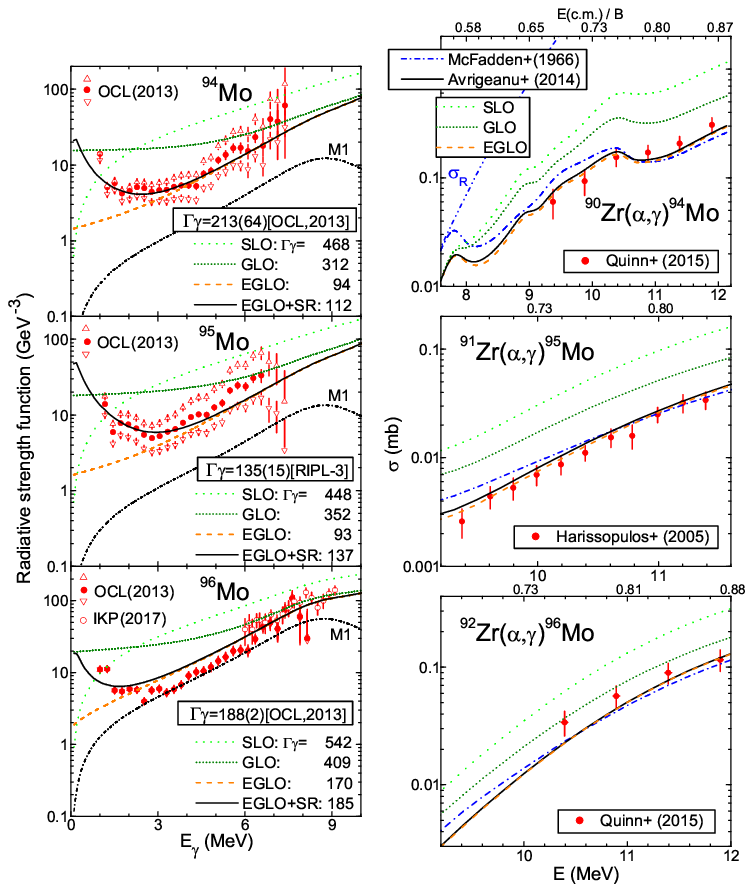}}
\caption{\label{fig:RSFag-Mo}(Color online) As Fig.~\ref{fig:RSFag-Ge} but for RSFs of the compound nuclei $^{94,95,96}$Mo \cite{hu13}, and $(\alpha,\gamma)$ reaction on $^{90,91,92}$Zr \cite{sjq15,sh05}.}
\end{figure*}

There are two different cases of the $(\alpha,\gamma)$ reaction cross sections below the Coulomb barrier $B$ \cite{wn80} which are illustrated in Fig.~\ref{fig:RSFag-Ge} for $^{70,74}$Ge target nuclei. First, the lower incident energies and the much larger threshold energy for the $^{70}$Ge$(\alpha,n)$$^{73}$Se reaction led to the total-reaction cross section going in the $\gamma$-channel at least in the first half of the incident-energy range of Ref. \cite{zf96}. Thus, the $(\alpha,\gamma)$ reaction analysis becomes a powerful tool for the study of the  $\alpha$-particle OMP, while the effects of various RSF models on the calculated cross sections are yet close to the error bars of the measured data \cite{zf96}. The factor of $\sim$2 for the overpredicted cross sections by the global potential of McFadden and Satchler \cite{mcf66} is finally obvious at lowest energies. The good agreement of the measured and presently calculated cross sections is similar to that found previously \cite{ma09} with a former version of the same OMP \cite{va14}. One may note in this respect that the rather significant replacement of an early E1 model used in \cite{ma09} has little effect on the calculated cross sections since, beyond the minor RSF effects shown in Fig.~\ref{fig:RSFag-Ge} for this reaction, the corresponding former RSF predictions were also checked versus the RSF and $\Gamma_{\gamma}$ data.

Second, higher incident energies of Ref. \cite{sjq15}, even if yet below $B$, as well as a threshold energy for $(\alpha,n)$ reaction on $^{74}$Ge which is nearly half of that for $^{70}$Ge, lead to $^{74}$Ge$(\alpha,\gamma)$$^{78}$Se reaction cross sections lower than $\sigma_R$ by more than two orders of magnitude. Under these conditions, differences given by use of the two $\alpha$-particle OMPs \cite{va14,mcf66} are close to the experimental errors (Fig.~\ref{fig:RSFag-Ge}). Actually, a lower slope of the excitation curve provided by McFadden--Satchler OMP  contributes to a crossover of the two curves  at $\sim$0.7$B$. On the other hand, the RSFs become now of first importance for the suitable account of the measured data by model calculations, which corresponds solely to the EGLO model. 
The SR addition has no effect while the use of the GLO and SLO models leads to increased cross sections but with rather different energy dependences. Thus, nearly twice cross sections are provided by the GLO formula at the lowest incident energies, with respect to the EGLO results, while they are closer at higher energies. On the contrary, the SLO model provides an additional increase to that of the GLO, but going from around 50\% to over 150\% with energy increase. 

\subsection{$^{90,91,92}$Zr$(\alpha,\gamma)$$^{94,95,96}$Mo} \label{Zr}

The rather detailed discussion on the RSFs of $^{90}$Zr in Sec.~\ref{RSF} has already taken into account the recent measurements for $^{92,94}$Mo nuclei \cite{gmt16,OCL}. 
A similar analysis for the compound nuclei $^{94,95,96}$Mo is shown in Fig.~\ref{fig:RSFag-Mo}. 
These results were obtained using the GDR parameters adopted by Guttormsen {\it et al.} \cite{mg05} and the value $T_f$=0.35 MeV found to describe particularly the RSF for $^{94}$Mo \cite{rs13}. Moreover, the SR parameters given above for $^{90}$Zr were also used with only a change of the peak cross section of 0.06 mb for $^{94,96}$Mo. There are thus well described (Fig.~\ref{fig:RSFag-Mo}) the recently measured RSFs \cite{hu13} which have just received an independent confirmation \cite{dm17}.

This analysis provides an additional support for the suitable description of the measured RSF data as well as the $\Gamma_{\gamma}$ measured or derived values \cite{hu13} only by the EGLO+SR model. The supplementary SR contribution provides the low-energy upbend but a reduced contribution to the corresponding $\Gamma_{\gamma}$ values of the even-even nuclei $^{94,96}$Mo. Nevertheless, there is a difference from $\sim$16 to $\sim$8\% of this contribution, which is well related \cite{gmt16} to the increased $S$ value for the heavier nucleus. At the same time, a larger contribution of $\sim$32\% for the odd-mass nucleus $^{95}$Mo  (Fig.~\ref{fig:RSFag-Mo}) has confirmed its dependence on the odd-even character already pointed out in Sec.~\ref{RSF}.

Actually, there is a former agreement between the measured $(\alpha,\gamma)$ reaction cross sections below $B$ for $^{91}$Zr \cite{sh05} and the calculated values using the $\alpha$-particle  OMP \cite{va14}, shown in Fig. 1 of Ref. \cite{va14b}. While the effects of neutron as well as $\alpha$-particle OMPs on the calculated cross sections were proved there to be similar to those of an EGLO model of the corresponding RSF, it is shown in Fig.~\ref{fig:RSFag-Mo} that much larger overpredictions follow the use of the GLO and especially SLO models. 

Similar results have been obtained in the case of the new data for $^{90,92}$Zr target nuclei \cite{sjq15}, with the only difference that the larger threshold energy of the $(\alpha,n)$ reaction on $^{90}$Zr makes possible the study of $(\alpha,\gamma)$ reaction cross sections closer to $\sigma_R$. Thus, just above the $(\alpha,n)$ threshold both SLO and GLO models overestimate by a factor of $\sim$2 the $\alpha$-particle capture cross section corresponding to the EGLO+SR model, with only a minor difference between them. However, at the higher energies of Ref. \cite{sjq15} there is a similar factor $\sim$2 between SLO and GLO related results, on the one hand, and between the GLO and either the EGLO+SR ones or the measured data, on the other. 

Nevertheless the new $(\alpha,\gamma)$ reaction data for $^{74}$Ge and $^{90,92}$Zr isotopes  are well described by the same $\alpha$-particle  OMP \cite{va14}, following the use of consistent sets of the rest of SM parameters, while their former analysis \cite{sjq15} reported three different OMPs providing their best description by TALYS calculations. Moreover, none of these potentials was the most-physical 3rd version of Demetriou {\it et al.} \cite{pd02} parameter sets, but the former two and the schematic initial approach in TALYS \cite{TALYS}. This fact proves the usefulness of  consistent SM parameter sets versus the attempts to determine which various parameter combination best describes the data. On the other hand, the present additional validation of the $\alpha$-particle OMP \cite{va14} in the incident channel for $A$$\sim$90 nuclei, represents a sound basis for a similar analysis of the $\alpha$-particle emission.

\section{Neutron-induced reactions on Zr stable isotopes} \label{Res}

\begin{figure*} %[t]
\resizebox{1.5\columnwidth}{!}{\includegraphics{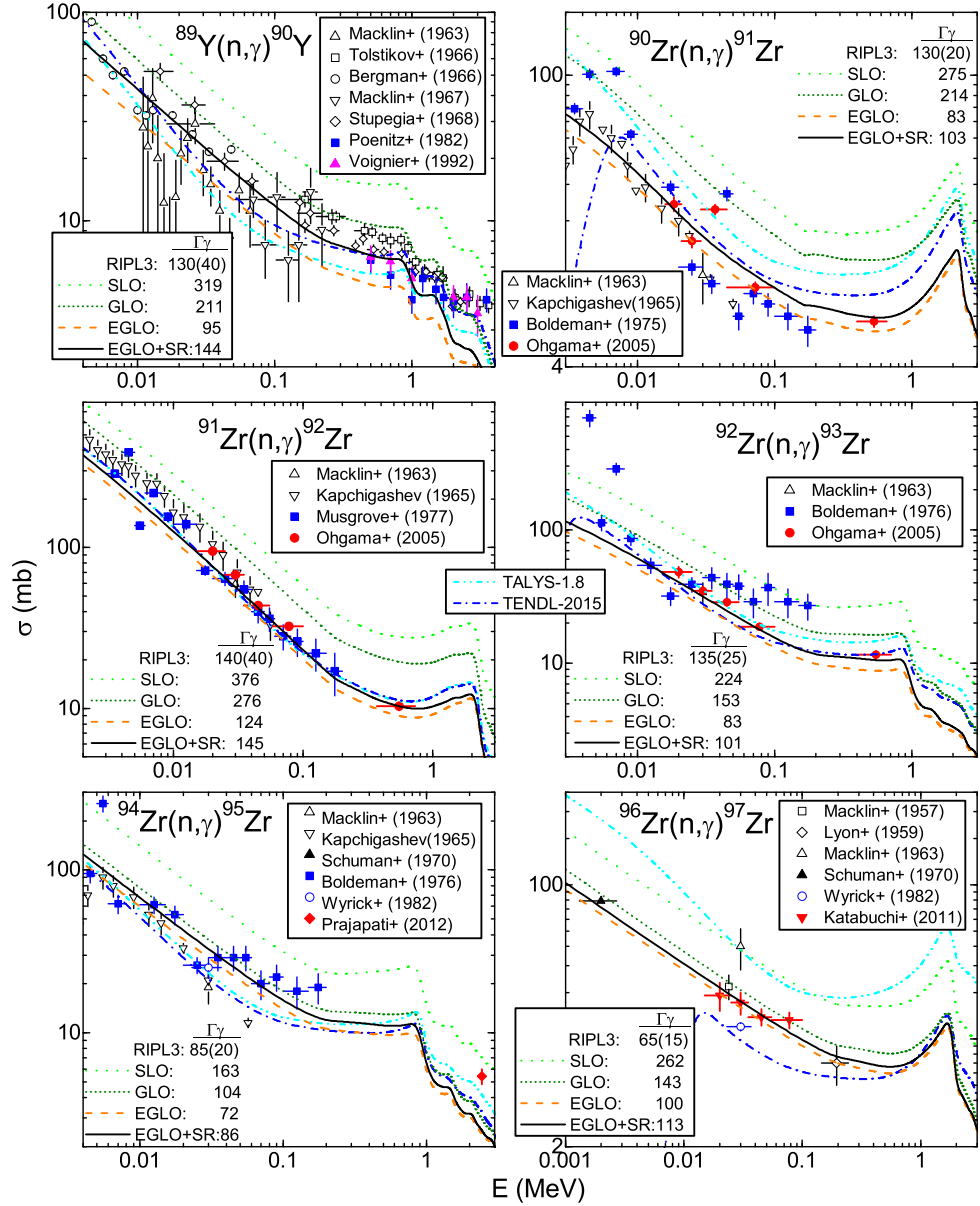}}
\caption{\label{fig:YZrng}(Color online) Comparison of neutron-capture cross sections for $^{89}$Y and $^{90-92,94,96}$Zr, measured \cite{exfor}, evaluated within the TENDL-2015 library \cite{TENDL15} (dash-dotted curves), and calculated using the code TALYS-1.8 \cite{TALYS} and its default parameters (dash-dot-dotted curves) as well as in the present work using the E1 radiations RSF models SLO (dotted curves),  GLO (short-dotted curves), EGLO (dashed curves), and EGLO+SR (solid curves). There are also shown the $s$-wave average radiation widths $\Gamma_{\gamma}$ (in meV) either measured \cite{ripl3} or corresponding to M1 and each of the above-mentioned E1 functions.}
\end{figure*}

\subsection{Neutron-capture systematic analysis} \label{CapRes}

The $(n,\gamma)$ reaction data analysis for the target nuclei $^{89}$Y and $^{90-92,94,96}$Zr (Fig.~\ref{fig:YZrng}) plays a similar role in check of the RSFs accuracy to that of the $(p,\gamma)$ reaction (Fig.~\ref{fig:SrYpg}).
The neutron energies considered in this respect were from above the resolved resonance range, as the assumption of an average statistical continuum overlap of available resonances becomes justified, and below 1-2 MeV. 
Thus, only the statistical decay of a CN in thermodynamic equilibrium contributes to the capture process, so that the comparison of the HF model calculations and measured cross sections provides a sound validation of the adopted RSFs.

The comparison of the experimental and calculated capture cross sections has pointed out again a good agreement for the use of the EGLO+SR models. 
One may note now that the SR contribution is rather small,  i.e. below 12\%, for $^{92}$Zr, while it increases for the odd nuclei $^{91,93,95,97}$Zr, and becomes largest for the odd-odd nucleus $^{90}$Y. Actually, the comparison of the measured \cite{ripl3} and calculated $\Gamma_{\gamma}$ values for the stable Zr isotopes (Fig.~\ref{fig:YZrng}) shows that the SR effect is just within the limit of the experimental error bars, while it may strongly affect the related quantities close to the neutron drip line \cite{sg15}.

An increase by even a factor $\sim$2 of the calculated capture cross-section is obtained once more if the EGLO+SR model is replaced by the GLO one. 
Moreover, a similar increase follows also the alternate use of the SLO model, with even larger values for $^{95,97}$Zr and similar effects also on the calculated $\Gamma_{\gamma}$ values. These changes, corroborated with the crucial role of RSF knowledge for the neutron capture account, are in agreement with the recent endorsement of the generally accepted validation of Hauser-Feshbach calculations, over an energy range of 0.01--10 MeV, within a factor of about 3 \cite{mb14}.

\begin{figure*} %[t]
\resizebox{1.5\columnwidth}{!}{\includegraphics{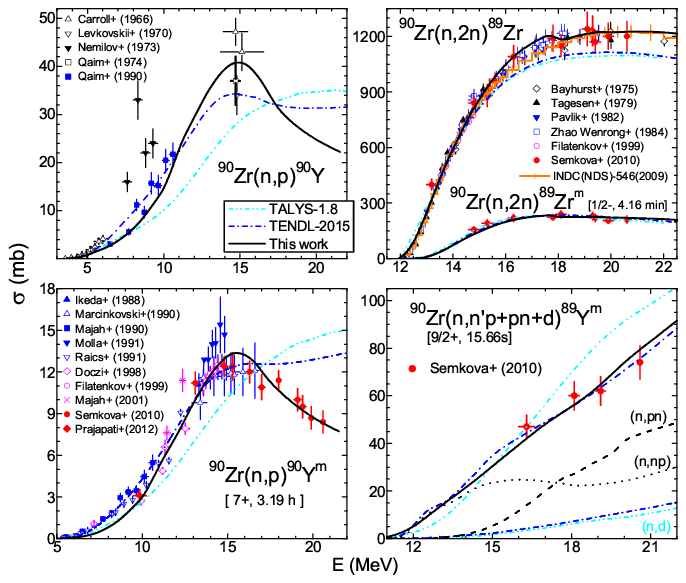}}
\caption{\label{fig:Zr90}(Color online) Comparison of measured \cite{vs10,exfor}, evaluated (dash-dotted curves \cite{TENDL15}, and histogram \cite{kiz09}), and calculated using the code TALYS-1.8 \cite{TALYS} and its default parameters (dash-dot-dotted curves) as well as in the present work (solid curves) fast-neutron reaction cross sections for $^{90}$Zr target nucleus. There are also shown the currently calculated excitation functions of the reactions $(n,n'p)$ (dotted curve) and $(n,pn)$ (dashed curve).}
\end{figure*}

\subsection{Systematic analysis of nucleon emission} \label{NemRes}

The first requirement of consistent nuclear-model calculations to use the same values for  parameters which are involved within various mechanisms made the object of Sec.~\ref{SMcalc}.
It should be followed by the suitable description of all available data for various reaction channels, which is proved hereafter for each of the stable Zr isotopes. However, the $\alpha$-particle emission data are discussed all together ultimately because of the more serious question marks on the related OMP.

\subsubsection{$^{90}$Zr} \label{90zr}

The previous detailed analysis of fast-neutron reactions on Zr isotopes \cite{vs10} concluded that the lack of agreement of the measured and calculated cross sections, especially for the $(n,p)$ reactions is due to PE effects. Therefore, our first interest in the present work concerned the PE suitable account. The particular semi-magic nucleus $^{90}$Zr could be most useful in this respect due to the related {\it $(n,2n)$ reaction} with more than only one measured data set above $\sim$15 MeV (Fig.~\ref{fig:Zr90}) as in the case of the other fast neutron-induced reactions on Zr isotopes. Thus, the model description of this $(n,2n)$ excitation function, with the maximum value larger than 1 b at an incident energy even above 20 MeV, stands for a sensitive check of the PE approach.

Actually there is a distinct shape of this  $(n,2n)$ excitation function for even-even target nuclei with a closed neutron shell as, e.g., $^{92}$Mo \cite{pr05}, making its discussion really interesting. Beyond the high threshold, following the corresponding large neutron $S$ value, its maximum of about 1.2 b is spread over $\sim$5 MeV. This unusual form is not fully described by neither the latest IAEA evaluation \cite{kiz09}, which underestimates the first quarter of this flat maximum, nor the TALYS-1.8 results and TENDL-2015 evaluated-data library which have a closer shape but prove an underestimation of $\sim$10\% (Fig.~\ref{fig:Zr90}). Its suitable account by the present calculations is merely due to the $l$-dependent PE modeling within the GDH model, namely the successive opening of various partial-wave contributions as it was discussed at large previously \cite{pr05,ma94}. Since the onset of these contributions is sharp within the GDH formalism including the advanced PLD \cite{ma98}, we smooth usually the related unphysical cross-section changes, over 1-2 MeV of the calculated excitation functions. However, we show now for this $(n,2n)$ reaction exactly the decrease given by the onset of the neutron PE contribution for the $l$$=$6$\hbar$ partial wave at the incident energy of $\sim$18 MeV (Fig.~\ref{fig:Zr90}). It is this onset supporting the enlarged maximum of the $(n,2n)$ reaction on $^{90}$Zr, in close agreement with the more recent measured data \cite{vs10}. Moreover, it could be underlined that the largest difference between our results and the IAEA evaluation \cite{kiz09} has been just before this point, making obvious the importance of a suitable PE account.

On the other hand, one may note that there is no change at the same energy of either the measured \cite{vs10} or the calculated $^{89}$Zr$^m$ isomeric cross sections (Fig.~\ref{fig:Zr90}). 
The lowest spin of the corresponding 1/2$^-$ state is related to the PE lack of importance for the low-spin states, which might also explain so close global \cite{TALYS,TENDL15} and present calculated results.

{\it The $(n,p)$ reaction} on the same target nucleus has the calculated excitation function (Fig.~\ref{fig:Zr90}) notably influenced by the onset of the proton PE contributions for the $l$$=$5 and 6 $\hbar$ partial waves at the incident energies of $\sim$10.4 and $\sim$20 MeV, respectively. The corresponding cross-section increases are important for its rising and especially decreasing sides, respectively. The good agreement between the measured \cite{vs10} and calculated cross sections for the high-spin 7$^+$ isomeric state has thus proved the PE suitable account, while no data exist above 15 MeV for the total $(n,p)$ reaction cross sections, with only one data set within the latest two decades and rather large spread of data at $\sim$14 MeV. The difference between the present work and TALYS-1.8 as well as TENDL results, close to a factor of 2 at higher incident energies, should be also due to the different PE models.
 
{\it The $(n,n'p+pn+d)$ reaction} leading to population of the larger-spin 9/2$^+$ isomeric state (Fig.~\ref{fig:Zr90}) has also the advantage of the measured data above 15 MeV, making possible an additional insight into the modeling suitability. Thus, the apparent change of the experimental excitation-function slope just above the incident energy of 18 MeV could be well related to (i) the slight decrease of the $(n,pn)$ reaction cross sections at this energy, followed by its slight increase at $\sim$20 MeV, due to the onset of PE contribution for 6 $\hbar$ partial wave of neutrons and protons, respectively, and (ii) the smaller weight of the $(n,n'p)$ reaction cross sections due to the dominant $(n,2n)$ reaction channel. The TALYS-1.8 calculation results for the $(n,d)$ reaction contribution to the same residual-nucleus population have been added to the sum of the just mentioned two-particle reaction channels, with no real effect on the excitation-function shape.

\subsubsection{$^{91}$Zr} \label{91zr}

\begin{figure} %[t]
\resizebox{0.75\columnwidth}{!}{\includegraphics{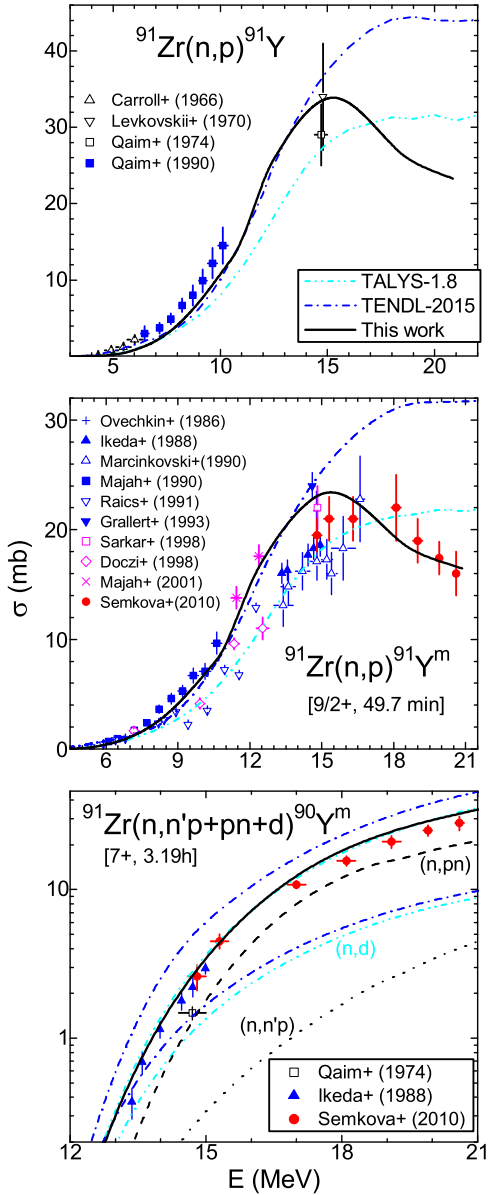}}
\caption{\label{fig:Zr91}(Color online) As Fig.~\ref{fig:Zr90} but for $^{91}$Zr \cite{vs10,exfor}.}
\end{figure}

{\it The $(n,p)$ reaction} data available so far are somehow parallel to the similar reaction on $^{90}$Zr, with a scarce body of total reaction cross sections up to $\sim$15 MeV, and more recent data even beyond 15 MeV \cite{vs10} for the $^{91}$Y$^m$ isomeric state with the  spin 9/2$^+$ (Fig.~\ref{fig:Zr91}). Thus the isotopic effect,  i.e. the decrease with $A$ of the $(n,p)$ reaction cross sections at the incident energy of $\sim$14 MeV for isotopes of the same element \cite{nim77}, is evident only for the total cross section while the larger isomeric cross sections may provide a better modeling check. Unfortunately, these isomeric data are rather scattered so that a TALYS-1.8 underestimation of them below 12 MeV led, as a result of these low-energy data account, to a TENDL-2015 overestimation by a factor of $\sim$2 for the data above 15 MeV.

\begin{figure*} %[b]
\resizebox{1.5\columnwidth}{!}{\includegraphics{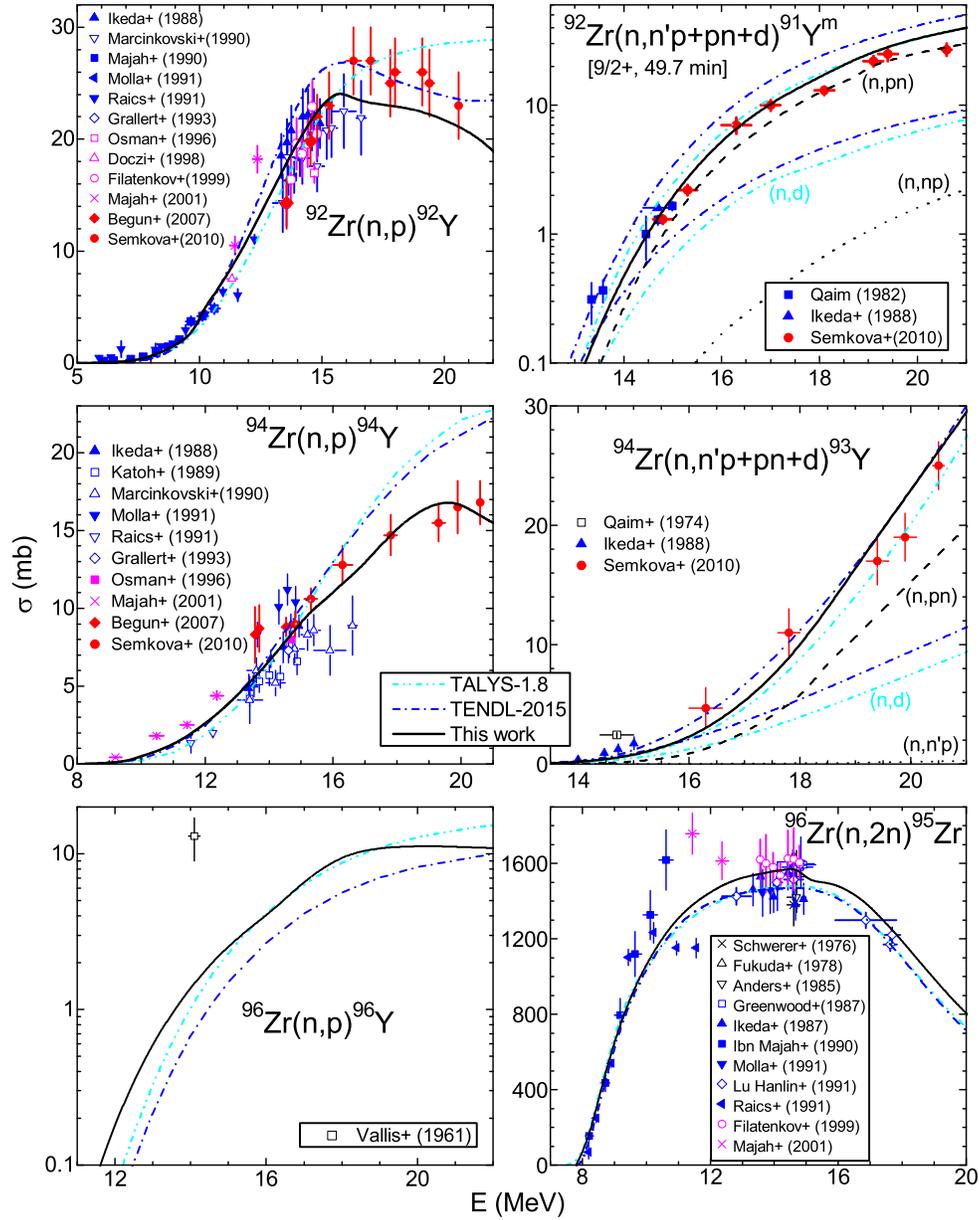}}
\caption{\label{fig:Zr9246}(Color online) As Fig.~\ref{fig:Zr90} but for $^{92,94,96}$Zr \cite{vs10,exfor}.}
\end{figure*}

The calculated $(n,p)$ reaction cross sections in the present work are characterized by the proton PE contribution onset for the $l$$=$5 and 6 $\hbar$ partial waves at the higher incident energies of $\sim$11.5 and $\sim$21.5 MeV, respectively. This accounts for the excitation-function rather sudden increase above 11 MeV, and the attenuated decrease from $\sim$19 MeV. The good agreement with the data above 15 MeV has been obtained at the same time with a similar one for the $(n,n'p+pn+d)$ reaction on $^{92}$Zr, leading to the population of the same 9/2$^+$ isomeric state of $^{91}$Y$^m$ to be discussed in Sec.~\ref{92zr} (Fig.~\ref{fig:Zr9246}).

{\it The $(n,n'p+pn+d)$ reaction} on $^{91}$Zr is followed, on the other hand, by population of the same 7$^+$ isomeric state $^{90}$Y$^m$ (Fig.~\ref{fig:Zr91}) as within the above-discussed $(n,p)$ reaction on $^{90}$Zr.
It should be also noted that the reaction $(n,pn)$ is the dominant component of the former sum. Therefore, the similar good agreement with the measured data also for the $(n,n'p+pn+d)$ reaction, which are less scattered and thus more confident, provides an additional support for the calculation results related to the $(n,p)$ reaction on $^{91}$Zr. This case could be compared to that of the TALYS-1.8 results which are very close to the present work for the $(n,n'p+pn+d)$ reaction but with the shape as well as absolute values at variance with the measured data.

\subsubsection{$^{92}$Zr} \label{92zr}

{\it The $(n,p)$ reaction} data available for this target nucleus (Fig.~\ref{fig:Zr9246}) are less scattered and thus more confident. On the other hand, the model calculations have to face the reduced existing knowledge of the low-lying levels for the more neutron-rich Y isotopes, thus identified also in Table 4 of Ref. \cite{ajk08}. Indeed, while there are obvious shell effects for $^{88,90}$Y, with only a neutron less or in addition to the closed shell $N$=50, even fewer levels are known for the heavier Y isotopes. 

Under these conditions, we have firstly considered the level-density parameter $a$-values provided by the smooth-curve method \cite{chj77} for $^{92,94,96}$Y, and obtained the back-shift $\Delta$-values by the fit of levels given in Table~\ref{densp} for  $^{92}$Y. Our main reason for the choice of this fit is that it corresponds also to the fit of the low-lying levels of $^{86}$Y (Table~\ref{densp}) which seems to be the Y isotope with slighter shell effects and closest to an accurate knowledge. The cumulative numbers of levels at the excitation energy of the highest excited level involved within present SM calculations (Table~\ref{densp}), corresponding to these BSFG parameters, are indeed rather double of those already known at the same energy \cite{ensdf}. However, one may note that even larger level densities are involved in the default TALYS-1.8 calculations at the lowest excitation energies, with the results also shown in Fig.~\ref{fig:Zr9246} in rather good agreement with the measured data up to $\sim$15 MeV.

Two effects related to the CN and PE mechanisms are present at energies above the maximum of this excitation function, in comparison with the lighter Zr isotopes. First, the CN contribution to this reaction cross section becomes smaller due to the $(n,p)$ reaction isotopic effect \cite{nim77}. Second, the onset of the proton PE contribution for $l$$=$6 $\hbar$ takes place at a lower incident energy, namely below 19 MeV. Together, these two effects provides a less decreasing high-energy side of the excitation function, at variance with the previous \cite{vs10} and actual TALYS-1.8 results which are still increasing with energy. The TENDL-2015 adjustment in this respect, taking into account the recent data \cite{vs10}, resulted however in a too constant shape up to $\sim$25 MeV, while the cross sections calculated in this work are in agreement only with the low limit of the data errors. The usefulness of further measurements above 20 MeV \cite{xl14} is thus obvious.

{\it The $(n,n'p)$ reaction} analysis, for the same isomeric state $^{91}$Y$^m$ populated also through the  $(n,p)$ reaction on $^{91}$Zr, is nevertheless helpful for the assessment of a suitable measured-data account by this work. Thus, the data are well described up to $\sim$17 MeV, while an apparent underestimation above this energy could be related to the underestimation of the  $(n,p)$ reaction data at the same energies. At the same time, the TENDL-2015 evaluation provides larger cross section for both reactions. 

\subsubsection{$^{94}$Zr} \label{94zr}

{\it The $(n,p)$ reaction} excitation function is marked by the same lower values of the CN contribution, while the onset of the proton PE contribution for $l$$=$6 $\hbar$ takes place at a similar incident energy. Consequently, its maximum is situated notably above 19 MeV. The recent data \cite{vs10} may suggest in this respect an energy even higher than 20 MeV, the agreement of the present calculations being only in the limit of the error bars (Fig.~\ref{fig:Zr9246}). 

{\it The $(n,n'p)$ reaction} complementary analysis has fortunately been of additional support of the presently calculated results. It is provided by the suitable account of both the energy dependence and cross-section values in Fig.~\ref{fig:Zr9246}, which are closely related to those of the $(n,p)$ reaction. 

\subsubsection{$^{96}$Zr} \label{96zr}

{\it The $(n,2n)$ reaction} is also of large interest for the present work, in spite of fewer measured data sets in comparison with the lightest stable isotope $^{90}$Zr, and particularly no more recent data towards the incident energy of 20 MeV. This is due to both its well increased cross sections with reference to $^{90}$Zr, following the similar isotopic effect but opposite for the $(n,2n)$ reaction, and excitation-function distinct shape, with the maximum at or even below 15 MeV. Thus, a simultaneous analysis of the  $(n,2n)$ reaction on the most neutron-poor and -reach Zr stable isotopes may provide a significant check of the PE description proved as a real modeling problem \cite{vs10}. 

Therefore, we show in Fig.~\ref{fig:Zr9246} exactly the decrease given by the onset of the neutron PE contribution for the $l$$=$6$\hbar$ partial wave at an incident energy of $\sim$15 MeV. Its place along the excitation function is rather opposite to that of the approaching the maximum for $^{90}$Zr (Fig.~\ref{fig:Zr90}), namely at the beginning of the decreasing side. However, the calculated results are in good agreement with measured data even in the present case, including a rather large decrease from the maximum at 13.5--15 MeV to the existing data at 17--18 MeV. Actually this decrease is well described particularly due to the noted PE increase.

\begin{figure*} %[t]
\resizebox{1.5\columnwidth}{!}{\includegraphics{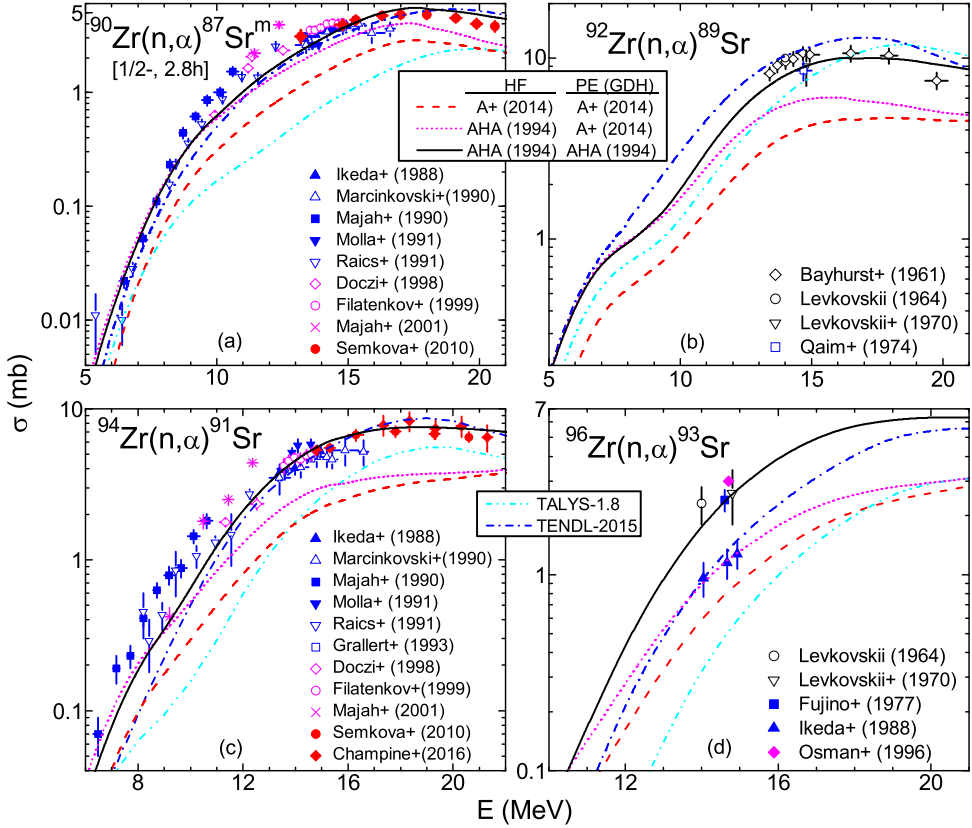}}
\caption{\label{fig:Zr90246na}(Color online) Comparison of $(n,\alpha)$ reaction cross sections for $^{90,92,94,96}$Zr target nuclei, measured \cite{bc16,vs10,exfor}, evaluated \cite{TENDL15} (dash-dotted curves), and calculated using the code TALYS-1.8 \cite{TALYS} and its default parameters (dash-dot-dotted curves) as well as in the present work using the $\alpha$-particle global OMPs of (i) Ref. \cite{va14} for the HF and PE model calculations (dashed curves), (ii) Ref. \cite{va14} only for the PE model while Ref. \cite{va94} is used for HF (short-dotted curves), and (iii) Ref. \cite{va94} for both HF and PE model calculations (solid curves).}
\end{figure*}

{\it The $(n,p)$ reaction} on the heaviest Zr stable isotope was, on the other hand, the object of only an early measurement around 14 MeV, with an overestimated value according to the isotopic effect. The calculated excitation function is shown in Fig.~\ref{fig:Zr9246} mainly to complete the related systematics for Zr isotopes.  
The proton PE contribution onset for the $l$$=$6 $\hbar$ takes place at a lower incident energy just above 17 MeV, leading to an excitation-function maximum similar to the case of $^{94}$Zr but broader due to a larger PE contribution.

\section{The $\alpha$-particle emission} \label{aOMP}

The analysis of the $(n,\alpha)$ reaction data available for Zr isotopes has the advantage of consistent data sets measured earlier on $^{90,94}$Zr at incident energies up to $\sim$15 MeV as well as the more recent to $\sim$21 MeV \cite{bc16,vs10}. On the other hand, there are only several either early or rather scattered data for $^{92,96}$Zr (Fig.~\ref{fig:Zr90246na}). Therefore we have paid full attention firstly to the former two isotopes, the corresponding conclusions being then considered for the other ones. 

\subsection{The $^{90,94}$Zr$(n,\alpha)$ reactions} \label{9094zr}

First, the recent OMP for $\alpha$-particles \cite{va14}, which provides a suitable description of the $\alpha$-particle induced reaction data within the wide mass range 45$\leq$$A$$\leq$209, was used within HF as well as PE calculations. Despite the good agreement obtained for a similar analysis of proton-induced reaction on Zn isotopes \cite{va15}, large discrepancies are obvious in Fig.~\ref{fig:Zr90246na}(a,c) between the measured and calculated cross sections for both isotopes $^{90,94}$Zr at all energies. 

Second, we replaced the above-mentioned $\alpha$-particle potential by the one found to describe well the $\alpha$-particle evaporation in neutron-induced reactions \cite{va94}, for calculation of $\alpha$-particle transmission coefficients involved in the HF calculations. The results were in much better agreement with the measured data particularly at incident energies below $\sim$9 MeV. 
It could be useful to note that the PE effects are yet rather low at these incident energies, while mainly the ground states of the residual nuclei are populated by $\alpha$-particles with  energies below 10 MeV.

Third, the same replacement concerned also the calculation of the corresponding PE intranuclear transition rates within the generalized GDH model. A rather good agreement has then been obtained also at the higher incident energies, where the residual nuclei are populated in continuum, close or beyond $S_n$, by $\alpha$-particles with energy centroids around 12-13 MeV. Actually, the PE weight within the $\alpha$-particle emission increases from $\sim$1 to $\sim$42\%, for  incident energies between 7--21 MeV on $^{90}$Zr, and from $\sim$5 to $\sim$84\% in the same energy range for $^{94}$Zr. The quite larger PE contribution for $^{94}$Zr follows the decrease of the CN component due to the isotopic effect of the also $(n,\alpha)$ reactions \cite{nim77,ma92}. 

A slight underestimation of data within a couple of MeV around the incident energy of $\sim$12 MeV could be due to a possible enhancement related to the position of a giant quadrupole resonance (GQR) in these nuclei, similar to the case of the Mo isotopes \cite{ma06}. 

The above final agreement should be considered at once with the basic differences between the imaginary-potential types and depths of the two $\alpha$-particle OMPs involved in this work. Thus, the  recent potential \cite{va14} concerns only surface absorption at lowest $\alpha$-particle energies, triggered by $\alpha$-induced reaction modeling. This surface term has firstly a constant depth $W_D$=4 MeV, then increasing for $\alpha$-particle energies from $\sim$8 to $\sim$12 MeV, followed by a decrease at once with the volume absorption depth increasing from $W_V$=0 at $\sim$10 MeV.  On the other side, the earlier and rather schematic potential \cite{va94} was obtained as an extrapolation to low energies of the global potential of Nolte {\it et al.} \cite{mn87} for  $\alpha$-particle energies above 80 MeV. Its imaginary component has only a volume depth increasing with energy from $W_V(0)$$\approx$5 MeV. 

Consequently, it may result first that a volume component is needed for an optical potential able to describe the $\alpha$-particle evaporation with the lowest energies, while only surface absorption matters in the incoming channel at the same energies. This standpoint is consistent with the volume multi--step interaction of neutrons with nuclei, followed by $\alpha$-particle emission from the same nuclear region, while the interaction of the incident $\alpha$-particles with similar energies takes place only within the nuclear surface (see, e.g., \cite{ma03} and Refs. therein). Second, this $\alpha$-emission OMP \cite{va94} seems to provide also a suitable PE description, in spite of an apparent inconsistency with the nuclear-surface PE localization in nucleon-induced reactions at low energies \cite{ma96}. It has indeed been found that even for a nuclear volume absorption, as it is the case of nucleons at incident energies of the order of the nuclear potential depth, the probability of the first nucleon-nucleon ($NN$) interaction along the trajectory of the projectile in pre-equilibrium reactions has its maximum within the nuclear surface (e.g., Fig. 1 of \cite{ma96}). However, the case of first interaction with an $\alpha$-particle has not yet been considered in a similar way. 

\subsection{The $^{92,96}$Zr$(n,\alpha)$ reactions} \label{9296zr}

The same assumptions have been involved also in the analysis of the corresponding data for $^{92,96}$Zr nuclei, with no further inference due to current level of these data. The related results shown in Fig.~\ref{fig:Zr90246na}(b,d) have led to the same conclusions as above. It may be useful to note that the smaller increase of the calculated excitation functions for $^{92}$Zr nucleus between 7--9 MeV follows the presence of only two excited levels up to an excitation energy of $\sim$1.9 MeV of the residual nucleus $^{89}$Sr, with only one neutron above the closed shell $N$=50.

An additional comment may concern the case of $(n,\alpha)$ reaction on $^{92}$Zr which was the object of a particular discussion with reference to the cross-section uncertainties due to the use of various level-density approaches (Fig. 9 of Koning {\it et al.} \cite{ajk08}), at the same time with a similar analysis for the neutron capture. Thus, changes of 25--33\% were found for the calculated $(n,\alpha)$ reaction cross sections around 14 MeV, due to different local and global NLD models, all of them being however larger than the measured data [Fig.~\ref{fig:Zr90246na}(b)] by $\geq$240\%. Once again, it results the key importance of the $\alpha$-particle OMP for the description of the $(n,\alpha)$ reactions which is one of great intricacy at incident energies where also the level densities and PE modeling cannot be disregarded. 

\subsection{The $^{91}$Zr$(n,\alpha$$n+n'\alpha)^{87}$Sr$^m$ reaction} \label{91zrna}

\begin{figure} %[t]
\resizebox{0.75\columnwidth}{!}{\includegraphics{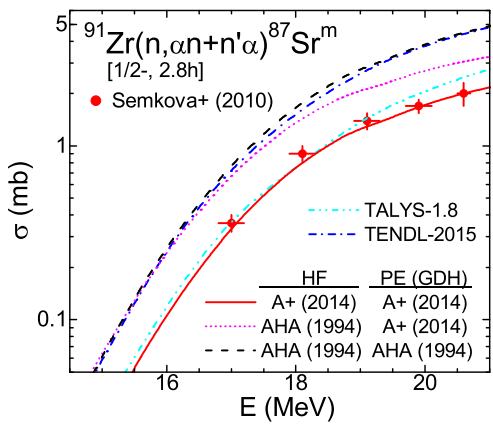}}
\caption{\label{fig:Zr91nan}(Color online) Comparison of $^{91}$Zr$(n,\alpha$$n+n'\alpha)^{87}$Sr$^m$ reaction reaction cross sections measured \cite{vs10}, evaluated \cite{TENDL15} (dash-dotted curves), and calculated using the code TALYS-1.8 \cite{TALYS} and its default parameters (dash-dot-dotted curves) as well as in the present work using the $\alpha$-particle global OMPs of (i) Ref. \cite{va14} for the HF and PE model calculations (solid curves), (ii) Ref. \cite{va14} only the PE model while Ref. \cite{va94} is used for HF (short-dotted curves), and (iii) Ref. \cite{va94} for both HF and PE model calculations (dashed curves).}
\end{figure}

There is, however, yet another related data set, namely for $^{91}$Zr$(n,\alpha$$n+n'\alpha)^{87}$Sr$^m$ reaction \cite{vs10}, which make it difficult to draw firm conclusions on the OMP suitable to account for the $\alpha$-particle emission. The same analysis as for the $(n,\alpha)$ reaction on the even-even Zr isotopes (Fig.~\ref{fig:Zr90246na}) has led to the opposed results shown in Fig.~\ref{fig:Zr91nan}. The agreement between the measured and calculated cross sections is provided by the $\alpha$-particle potential for incident $\alpha$-particles \cite{va14}. There are several points which should be considered in this respect, within the incident-energy range 17--21 MeV of the measured data \cite{vs10}.

(i) The contribution of the $(n,\alpha$$n)$ reaction to the activation of the isomeric state $^{87}$Sr$^m$ through $^{91}$Zr$(n,\alpha$$n+n'\alpha)^{87}$Sr$^m$ reaction is between 89--92\%. 

(ii) The weight of the $(n,\alpha$$n)$ reaction to the decay of the excited nucleus $^{92}$Zr by $\alpha$-particle emission is between $\sim$18--52\%.

(iii) The PE weight to the decay of the excited nucleus $^{92}$Zr by $\alpha$-particle emission is between $\sim$44--54\%, while the corresponding faster $\alpha$-particles populates mainly the residual nucleus $^{88}$Sr.

(iv) The activation cross sections for the isomeric state $^{87}$Sr$^m$ are around 16\% of the corresponding cross sections for $^{87}$Sr activation.

Therefore the minor $\alpha$-emission contribution to the activation of the isomeric state $^{87}$Sr$^m$ corresponds mainly to the lower-energy $\alpha$-particles emitted after the CN equilibration. They may be rather similar to the low-energy incident $\alpha$-particles, which are well described by the OMP of Ref. \cite{va14}, provided that their emission takes place however within nuclear surface.

\section{Conclusions} \label{Conc}

Recent accurate $(\alpha,\gamma)$ and $(n,\alpha)$ reaction data for Zr isotopes \cite{sjq15,bc16} as well as open questions pointed out by up-to-date systematic measurements and analysis of fast-neutron reactions on Zr isotopes \cite{vs10} have been considered in order to investigate the reliability of 
using a previous $\alpha$-particle global potential \cite{va14} for incident as well as emitted $\alpha$-particle model predictions. 
A consistent parameter set has been involved in this respect, established or validated by independent analysis of various experimental data \cite{eda80} other than the activation cross sections making the object of this work. Thus, the transmission coefficients of protons and $\gamma$ rays, given by the corresponding optical potential and $\gamma$-ray strength functions, respectively, have especially been fixed by independent analysis of $(p,n)$ reaction and radiative strength functions data, and then also checked by study of $(p,\gamma)$, $(\alpha,\gamma)$, and $(n,\gamma)$ reactions. Actually, the present work shows more exactly than formerly \cite{va15,va16} that the low-energy RSF enhancement in addition to the EGLO model does affect the calculated capture cross sections but much less than the use of either SLO or even GLO models. 

It has thus been possible to describe the new $(\alpha,\gamma)$ data as well as most of the available neutron-activation data for all Zr stable isotopes at once, with no empirical rescaling factors of the $\gamma$ and/or nucleon widths.
The usefulness of a consistent input parameters in nuclear model calculations is therefore proved, against the trial of different combinations of SM ingredients \cite{sg03,pt04} even self-consistent in the sense that they combine only phenomenological or microscopic models  \cite{sh13}. 
Moreover, this work has shown a definite proof of the pre-equilibrium emission description by the GDH model \cite{mb83} using advanced particle-hole level densities with a Fermi-gas energy dependence of the single-particle level density \cite{ck85}.
Thus, the successive opening of various partial-wave PE contributions to the $(n,p)$ and $(n,2n)$ reaction cross sections is finally leading to a suitable account of various excitation functions, including the changes from one isotope to another.

Finally, a still open question concerns the optical potential which may be able to describe the $\alpha$-particle emission at least in fast-neutron induced reactions. The present analysis of the $(n,\alpha)$ reaction on $^{90,94}$Zr pointed out that the $\alpha$-particle OMP \cite{va14}, which provides a suitable description of the $\alpha$-particle induced reaction data within the wide mass range 45$\leq$$A$$\leq$209, including Sr isotopes \cite{ao16}, has led to underestimated HF as well as PE calculation results. Much improved calculated cross sections which are obtained using the OMP found earlier to describe the neutron-induced $\alpha$ emission \cite{va94} make apparent two points. First, it may result that a volume component is needed for an optical potential able to describe the $\alpha$-particle equilibrium emission at so low energies that only surface absorption takes place in the incoming channel. Second, this OMP component is able to provide suitable PE description, in spite of an apparent inconsistency with the nuclear-surface PE localization in nucleon-induced reactions at low energies \cite{ma96}. On the other hand, validation of the former OMP \cite{va14} for the $\alpha$-particle emission in low-energy proton-induced reactions on Zn isotopes \cite{va15} could be related to a surface character of these reactions too, at possible variance to the fast-neutron induced reactions except the  $^{91}$Zr$(n,\alpha$$n+n'\alpha)^{87}$Sr$^m$ reaction data \cite{vs10}. Further measurements and analysis of $\alpha$-particle emission in neutron- as well as low-energy proton-induced reactions, in addition to related $(n,\alpha)$ and $(n,\alpha$$n)$ reactions on the same target nucleus, could make clear these points. 

\section{Acknowledgments}

This work has been partly supported by Autoritatea Nationala pentru Cercetare Stiintifica (PN-16420102) and partly carried out within the framework of the EUROfusion Consortium and has received funding from the Euratom research and training programme 2014-2018 under grant agreement No 633053. The views and opinions expressed herein do not necessarily reflect those of the European Commission.

\end{document}